# Early Beam Injection Scheme for the Fermilab Booster: A Path for Intensity Upgrade

C. M. Bhat[1]

*Fermi National Accelerator Laboratory*
*P.O. Box 500, Batavia, IL, 60510, USA*

Over the past decade, Fermilab has focused efforts on the *intensity frontier* physics and is committed to increase the average beam power delivered to the neutrino and muon programs substantially. Many upgrades to the existing injector accelerators, namely, the current 400 MeV LINAC and the Booster, are in progress under the Proton Improvement Plan (PIP). Proton Improvement Plan-II (PIP-II) proposes to replace the existing 400 MeV LINAC by a new 800 MeV LINAC, as an injector to the Booster which will increase Booster output power by nearly a factor of two from the PIP design value by the end of its completion. In any case, the Fermilab Booster is going to play a very significant role for nearly next two decades. In this context, I have developed and investigated a new beam injection scheme called "early injection scheme" (EIS) for the Booster with the goal to significantly increase the beam intensity output from the Booster thereby increasing the beam power to the HEP experiments even before PIP-II era. The scheme, if implemented, will also help improve the slip-stacking efficiency in the MI/RR. Here I present results from recent simulations, beam studies, current status and future plans for the new scheme.

PRESENTED AT

DPF 2015
The Meeting of the American Physical Society
Division of Particles and Fields
Ann Arbor, Michigan, August 4-8, 2015

---

[1] Work supported by Fermi Research Alliance, LLC under Contract No. De-AC02-07CH11359 with the United States Department of Energy

## 1. OVERVIEW

Fermilab has been the US premier high energy physics (HEP) laboratory for the past four decades. Around 2000, alongside ppbar luminosity upgrade for the Tevatron at 2 TeV, the Fermilab Long Range Accelerator Program Planning started focusing on increasing the beam power on targets for neutrino beams. As the energy frontier shifted to the LHC at CERN, Geneva, Switzerland, Fermilab started improving its accelerator complex to establish a world-leading facility for particle physics research based on intense proton beams and to address many unsolved problems in the neutrino sector and rare processes and, possibly, physics beyond the *standard model* of the particle physics.

An aerial view of the Fermilab accelerator complex with current and future fixed target HEP experiments are shown in Fig. 1 (left). To increase the beam power on target we have undertaken a staged approach, 1) Proton Improvement Plan (PIP) [1] that helps delivering 700 kW beam on the NOvA target by the end of 2017 and 2) PIP-II [2] which adds a new 800 MeV LINAC to the Fermilab complex that uses superconducting RF technology and injects brighter beam than in the PIP era into the existing 8 GeV Booster. The beam power on target is expected to exceed a MW during the PIP-II era. Nevertheless, the current Booster would play a very significant role during PIP and PIP-II eras.

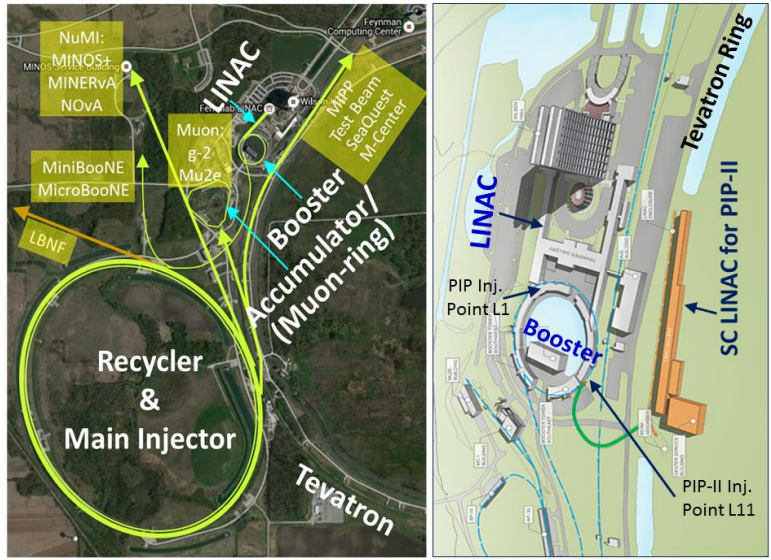

Figure 1: Schematic of the Fermilab complex with various ongoing and future HEP experimental programs (left). PIP and the proposed PIP-II (right). Beam injection points from the LINAC and the PIP-II LINAC are also shown.

The Fermilab Booster is the 2$^{nd}$ oldest rapid cycling proton synchrotron in the world (operating since 1971 [3]) with beam acceleration on a 15 Hz sinusoidal magnetic ramp. Initially, the proton beam at injection was at a kinetic energy of 200 MeV. Since 1978, Booster has operated with multi-turn $H^-$ charge exchange beam injection and takes about 2-40 µsec for injection (beam revolution period ≈2.21 µsec at injection). In 1990, the injection energy was upgraded to 400 MeV. Ever since the Booster came into operation, the beam injection was done always very close to $B_{min}$ (minimum of the magnetic field ramp). The Booster has, so far, operated with an average beam delivery cycle rate <15 Hz, much less than its design repetition rate. However, by the end of the completion of PIP the Booster will be capable of delivering beam at 15 Hz. During the PIP-II era the beam delivery rate from the Booster will also be increased from 15 Hz to 20 Hz. Thus, the full potential of the Booster is yet to be realized.

Recently, we proposed a new injection scheme [4], which fits well between PIP and PIP-II eras and has high potential to increase the beam power significantly. This paper presents a fully developed scheme called *Early Beam*

*Injection scheme* (EIS), which involves beam injection on the deceleration part of the magnet ramp in the Booster. Here we explain the general principle of the method, the results from beam dynamics simulations and demonstration with beam experiments. Beam dynamics simulations applied to the EIS convincingly shows that the new scheme has many advantages over the one currently in use (referred to as the *old scheme*).

## 2. PRINCIPLE OF THE EARLY INJECTION SCHEME AND SIMULATIONS

Figure 2 shows schematic views of both the old beam injection scheme and the EIS. Historically, after the beam injection in to the Booster, the beam was allowed to debunch fully before the start of the beam capture, because the rf frequencies of the LINAC (≈200 MHz) and that of the Booster (≈37 MHz) did not match. In the old scheme the beam is already in the increasing B field of the main Booster dipoles even during debunching. Consequently, the beam capture and turn on of the beam acceleration was done as swiftly as possible; hence the emittance dilution and beam loss in the early part of the beam cycle was inevitable. On the other hand, the principle of the new scheme, the EIS, is quite different and is as follows: 1) beam is injected ≈150μsec before the $B_{min}$, 2) beam capture begins immediately after the injection, 3) $dP/dt = 0$ is imposed during beam capture although $B$ is changing, 4) changing $B$ field at a constant momentum still introduces varying revolution frequency in accordance with $\Delta B/B = \gamma_T^2 \Delta f/f$, this must be accounted for, where $\gamma_T = 5.478$ is transition gamma for the Booster. Since there is an ample time to make the beam capture as adiabatic as needed, beam loss and emittance growth can be eliminated in the early part of the cycle and for the rest of the acceleration cycle. The time required for the beam capture and the optimum rf voltage curve for the entire acceleration cycle are determined by longitudinal beam dynamics simulations using ESME [5]. These simulations use the measured beam energy spread (≈1.50 MeV, full energy spread [6]). Figure 3 shows the predicted phase space distributions and the corresponding line-charge distributions at various stages of the beam acceleration in the EIS. These simulations clearly display many benefits of the EIS over the old scheme - i) less than 5% longitudinal emittance dilution at capture with no losses (old scheme showed ≈50% emittance dilution with a few percent beam loss), ii) no further emittance dilution until transition crossing, iii) an additional rf phase shift of ≈6⁰ after transition phase jump from $\phi$ to $\pi$-$\phi$ would make a better match between the rf bucket and bunch distribution through the rest of the cycle. iv) A snap bunch rotation gives a final energy spread of $(\Delta E)_{RMS} \approx 2$ MeV for the beam bunches at extraction which is 30% less than that from the old scheme; overall emittance dilution in EIS is about 50% with no particle loss through the cycle. Beam with a smaller energy spread at extraction helps to improve slip stacking efficiency in the Main Injector/Recycle Ring [7]. v) Improved emittance preservation in the EIS implies reduced rf power by about 30% over the cycle. vi) Since there is more room for beam injection in the Booster one can easily accommodate a longer LINAC pulses than that used in the old scheme. Thus, one can increase the extracted beam intensity substantially from the Booster over the current operational value of about 5.5E10p per bunch, i.e., 4.3E12 p per Booster cycle. Simulations show a gain in beam intensity by more than a factor of two.

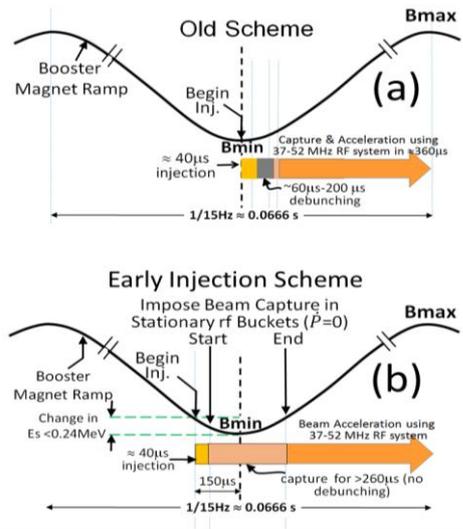

Figure 2: Schematics of old injection scheme and the EIS.

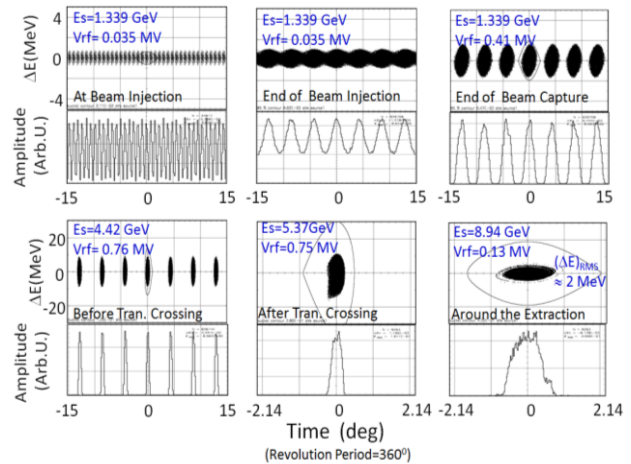

Figure 3: ESME [6] simulations for the EIS; (ΔE,Δt)-phase space distribution and line charge distributions at various instances in the Booster cycle.

## 3. BEAM EXPERIMENTS

Beam experiments have been carried out in two steps. First, we measured the energy acceptance of the Booster at 400 MeV by injecting ~1E12 p/batch at various times prior to $B_{min}$ with a fixed $B_{min}$. These measurements showed that one can inject beam without any loss up to ~500 μs prior to the $B_{min}$, implying that the total energy acceptance of the Booster is about 4.2±0.2 MeV which is larger than the beam energy spread of 1.5 MeV at injection [5]. However, during the bunching process of beam with the Booster rf system at 37 MHz the full energy spread of the beam increases from about 1.5 MeV to 3.5 MeV which is small enough to eliminate the beam losses at the early part of the beam cycle.

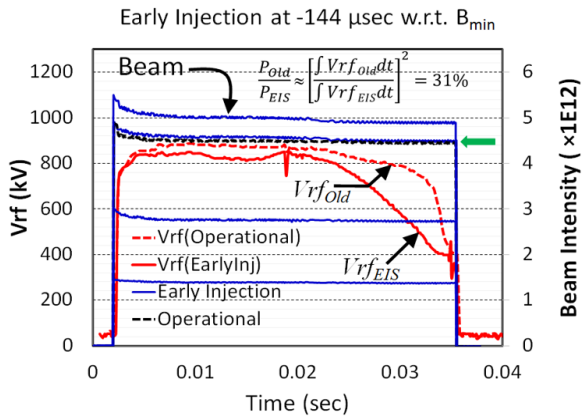

Figure 4: RF voltage (red) and beam (blue) through the Booster cycle for the EIS for four different initial beam intensities. Data for 4.8E12p/batch are shown for old-scheme (dashed curve) and the EIS for comparison.

Table I: PIP and PIP-II parameters. Numbers in bold are those with gains from the EIS scheme in the Booster.

| Parameters | PIP | PIP-II |
|---|---|---|
| Inj. Energy (KE) | 0.4 GeV | 0.8 GeV |
| Energy@Exit (KE) | 8 GeV | 8 GeV |
| Booster Repetition Rate | 15 Hz | 20 Hz |
| Booster Harmonic Number | 84 | 84 |
| Length of LINAC beam pulse | ≈30 (**42**) μs | ≈600 μs |
| Intensity@Inj. [p/batch] | 4.52E12 (×**1.4**) | 6.63E12 |
| Intensity@Exit [p/batch] | 4.3E12(×**1.4**) | 6.44E12 |
| Efficiency | 95 (**97**) % | 97 % |
| Booster Beam Power @Exit | 94 (**~130**) kW | 184 kW |
| NOvA Beam Power | 700 (**~950**) kW | 1.2 MW |

Finally, the beam experiments on the entire acceleration cycle have been carried out using the EIS following guidance from ESME simulations. The simulations indicated two possible modes for EIS from injection to the end of beam capture w.r.t. $B_{min}$, a) symmetrical and b) nonsymmetrical. We find that there was no advantage of one over the other scenario if beam is injected at about ≈-150 μsec w.r.t. $B_{min}$. Therefore, case "a" is chosen. Figure 4 shows the measured beam transmission and the rf voltages from injection to beam extraction in the Booster for the EIS for different beam intensities. For an injection intensity of 4.8E12p/batch, the old scheme (dashed curves) is compared with the EIS.

There were two important beam controls, critical for the full demonstration and implementation of the EIS, that were not available during the afore mentioned beam experiments. The EIS needed full control of rf frequency synchronization with changing magnetic field during the beam capture. The existing Booster LLRF control system, however, did not allow frequency control until about 10 μs after the $B_{min}$. Therefore, beam capture was started only after $B_{min}$ and hence, was not quite adiabatic. Secondly, the additional phase shift of $6^0$ after the transition crossing that required for better match between bunch shapes to bucket shapes was not available. As a result of this the displayed data does not show much improvement in the transmission efficiency over the old scheme. However, these experiments clearly showed that one can reduce the required rf power by nearly 30% for the EIS relative to the old scheme.

Table I summarizes the PIP and PIP-II performance goals. It also shows our expectation if the EIS is adopted during or after the completions of PIP. The LINAC is capable of providing >60 μs long $H^-$ pulse for injection [7] though we are using only less than half of it in current operation. Conservatively, one can expect an increase in beam intensity in the Booster with the EIS by at least 40% over the PIP design. This gives an output beam power of about 130 kW from the Booster and approximately 950 kW on the NOvA target.

As the beam intensity increases one must address intensity related issues such as beam loading compensation, coupled bunch instabilities and impedance issues in the Booster which are essential for the success of PIP-II. We might as well start the R&D related to high intensity beam as soon as the EIS is implemented in operation. Many of the needed improvements for implementing the EIS in operation are in progress.

## Acknowledgments

Author would like to thank W. Pellico, C. Drennan, F. Garcia, K. Triplett, S. Chaurize and T. Sullivan for their help during the various stages of beam measurements and Pushpa Bhat for many useful comments. Special thanks are due to B. Hendrik for his help in developing beam control software.